\pdfoutput=1
\documentclass[aps,amsfonts,pre,twocolumn,superscriptaddress,showpacs]{revtex4-1}
\usepackage{epsfig,amsmath,amssymb,bm,epsf,graphics,psfrag,verbatim,subfigure}

\usepackage[normalem]{ulem}
\usepackage{float}
\usepackage[usenames]{color}
\usepackage{graphicx}
\usepackage{ulem}
\usepackage{amsfonts}
\usepackage{amsmath}
\usepackage{natbib}
\usepackage{epsfig,amsmath,amssymb,bm,epsf,graphics,psfrag,verbatim,subfigure}

\begin{document}

\title{Modeling delayed processes in biological systems}
\author{Jingchen Feng}
\affiliation{Department of Bioengineering and Center for Theoretical Biological Physics, \\Rice University, Houston TX, 77251-1892}
\author{Stuart Sevier} 
\affiliation{Department of Physics and Center for Theoretical Biological Physics, Rice University, Houston TX, 77251-1892}
\author{Bin Huang}
\affiliation{Department of Chemistry and Center for Theoretical Biological Physics, Rice University, Houston TX, 77251-1892}
\author{Dongya Jia}
\affiliation{Graduate Program in Systems, Synthetic and Physical Biology and Center for Theoretical Biological Physics, Rice University, Houston TX, 77251-1892}
\author{Herbert Levine}
\affiliation{Department of Bioengineering and Center for Theoretical Biological Physics, \\Rice University, Houston TX, 77251-1892}
\date{\today}
\begin{abstract}
Delayed processes are ubiquitous in biological systems and are often
characterized by delay differential equations (DDEs) and their extension to include stochastic effects. DDEs do not explicitly
incorporate intermediate states associated with a delayed process
but instead use an estimated average delay time. In an effort to examine the validity of this approach,
we study systems with significant delays by explicitly incorporating
intermediate steps. We show by that such explicit models often yield significantly
different equilibrium distributions and transition times as compared
to DDEs with deterministic delay values. Additionally, different
explicit models with qualitatively different dynamics can give rise
to the same DDEs revealing important ambiguities. We also show that DDE-based predictions of oscillatory behavior may fail for the corresponding explicit model. 
\end{abstract}


\maketitle

\section{Introduction}

Delayed reactions are present in many biological systems. Most notably,
the central dogma of biology describes how functional protein production results from a sequence of
of numerous processes covering transcription, translation and post-translational modifications. The sequential nature of protein production
causes delay from the point that RNA polymerase binds to promoter DNA to
the appearance of fully functional proteins\cite{key-1,key-6,key-8}.
Moreover, the degradation of proteins can also require multiple steps\cite{key-7}.
In addition to delay created through reaction chains,
the transportation of molecules within a cell is a highly stochastic
diffusion process which itself can often generate significant delays within
a system. For example, in a eukaryotic cell mRNA is first produced in the nucleus and then
transported to the cytoplasm for further translation. Transportation can be viewed as a reaction chain if molecules
at different spatial points are treated as intermediate products.
However, the intermediate steps in the transportation process
are reversible (i.e. molecules are free to move back and forth); in
contrast, many reactions in protein production proceed in an irreversible
manner. In this paper, we focus on the later case and leave the
former case for future study. 

To date, delay in biological systems has been most
extensively studied through Delay Differential Equations (DDEs) and their extension to include stochasticity.
DDEs omit intermediate steps associated
with a delayed process and instead estimate the average
delay time for those steps. Typically fixed delay values are considered \cite{key-2, key-23, key-4, key-24, key-25, key-26, key-27}, though DDEs with a distribution of delay values have been studied \cite{key-21, key-22}. Several studies employ
DDEs to illustrate that delay can induce oscillation in otherwise
stable systems \cite{key-1,key-3,key-4,key-10,key-11,key-12,key-13,key-14}.
Intuitively, if we increase the delay from zero to a value comparable
to the residence time \cite{key-20} of the system, oscillations
may appear because of a phase lag in regulation. Additionally,
a recent study employing DDEs presented a less intuitive observation that a relatively
small transcriptional delay can stabilize bistable gene networks \cite{key-2}.
These studies demonstrate that a delay can greatly influence the dynamics and
equilibrium properties of biological systems. 

An obvious check on the validity of DDEs is to
compare them to more complete models that explicitly incorporate intermediate
steps into the system. We will refer to such models
as explicit models. In this study we compare the
predictions of fixed delay systems and explicit models. Instead of applying
delay differential equations  \cite{key-13,key-14}, we 
simulate reactions as delayed stochastic systems (DSSs)
using a Gillespie algorithm first proposed by Bratsun et al \cite{key-3}.
We show by a series of paradigmatic examples that DSSs with fixed delay often mischaracterize system behavior.  Our results should inject a needed note of caution into this common practice. 

The organization of the paper is as follows.
In section II, we discuss a self-activation circuit that has two stable
states, first studied as a delayed stochastic system 
by Gupta et al \cite{key-2}. We construct two distinct explicit
models for the same DSS and demonstrate that one model produces
results consistent with the DSS while the other produces markedly different
results. In section III, we discuss how the original DSS can sometimes emerge as the limit of an explicit model with many intermediate steps of equal mean duration. In section IV, we examine a toggle switch circuit, another common bistable system. In this case, we examine
an explicit model that again exhibits quantitatively different behavior
as compared to the parent DSS. In section V, we discuss
a simple linear system where a DSS with deterministic delay generates oscillations when explicit
models do not. In section VI, we summarize our work and its implications for constructing biological models.

\section{Self-activation circuit}
\subsection{Delay differential equations}

Consider the single-gene delayed positive feedback loop shown in Fig.1a. The
dynamic behavior of the average number of molecule
X is denoted by $x$ and is determined by the following DDE 
\begin{equation}
\dot{x}=\alpha+\beta\frac{x(t-\tau)^{b}}{c^{b}+x(t-\tau)^{b}}-\gamma x\label{eq:dde1}
\end{equation}
where $\alpha$ is the basal transcription rate due to leakiness of
the promoter, $\beta$ the increase in transcription rate due to protein
binding to the promoter, $b$ the Hill coefficient, $c$ the concentration
of $x$ needed for half-maximal induction, $\gamma$ the degradation
rate coefficient of the protein, and $\tau$ the transcriptional delay
time. With the parameter values used in \cite{key-2}, the self-activation
circuit is bistable. 

We are interested in the stochastic version of this type of delayed system. Here, the right hand side of (\ref{eq:dde1}) is re-interpreted as the rate for a reaction that produces an additional X. We employ the modified Gillespie algorithm first
proposed by Bratsun et al\cite{key-3} to carry out stochastic simulations.
Here are the formal steps:
\begin{enumerate}
\item Set initial states $\boldsymbol{X}=(X_{1},...,X_{N})$, set time $t=0$
and reaction counter i=1.
\item Calculate the rates of each reaction $a_{\mu},\mu=1,...,M.$
\item Generate two uniform random numbers $u_{1},u_{2}\in[0,1]$ 
\item Compute $\Delta t_{i}=-\ln(u_{1})/\sum_{\mu}a_{\mu}.$ The next reaction
is scheduled at $t+\Delta t_{i}$.
\item If there are delayed reactions scheduled within time interval $[t,t+\Delta t_{i}]$,
then step 2-4 are ignored. Update $t$ to the next scheduled delay
reaction time $t_{d}$. $\boldsymbol{X}$ states are updated according
to the delayed reaction channel, and update $i=i+1$. Go to step 2.
Otherwise, proceed to step 6. 
\item Find the channel of the next reaction $\mu$, namely take $\mu$ to
be integer for which $\sum_{j=1}^{\mu-1}a_{j}<u_{2}a_{t}\le\sum_{j=1}^{\mu}a_{j}$,
where $a_{t}=\sum_{j=1}^{M}a_{j}$ is the total rate. Update $t=t+\Delta t_{i}$. 
\item If the selected reaction $\mu$ is not delayed, update $\boldsymbol{X}$
according to the reaction channel, update
$i=i+1$. If the selected reaction is delayed, update is put off till
$t_{d}=t+\tau$. Go to step 2. 
\end{enumerate}

Results for the self-activation circuit from stochastic DDEs are shown in Fig.1b\&1c. (We have validated our results
by employing a different delay stochastic simulation method \cite{key-17,key-18}.)
When the system has instant feedback (zero delay), the equilibrium
distribution favors the low number state (Fig 1c) while for increasing
delay the high number state becomes more occupied. In addition, the
mean residence time (MRT), sometimes called the average first passage/transition
time, of the low number state grows rapidly with increasing delay. 
\begin{figure*}
\centering
\includegraphics[width=0.80\textwidth]{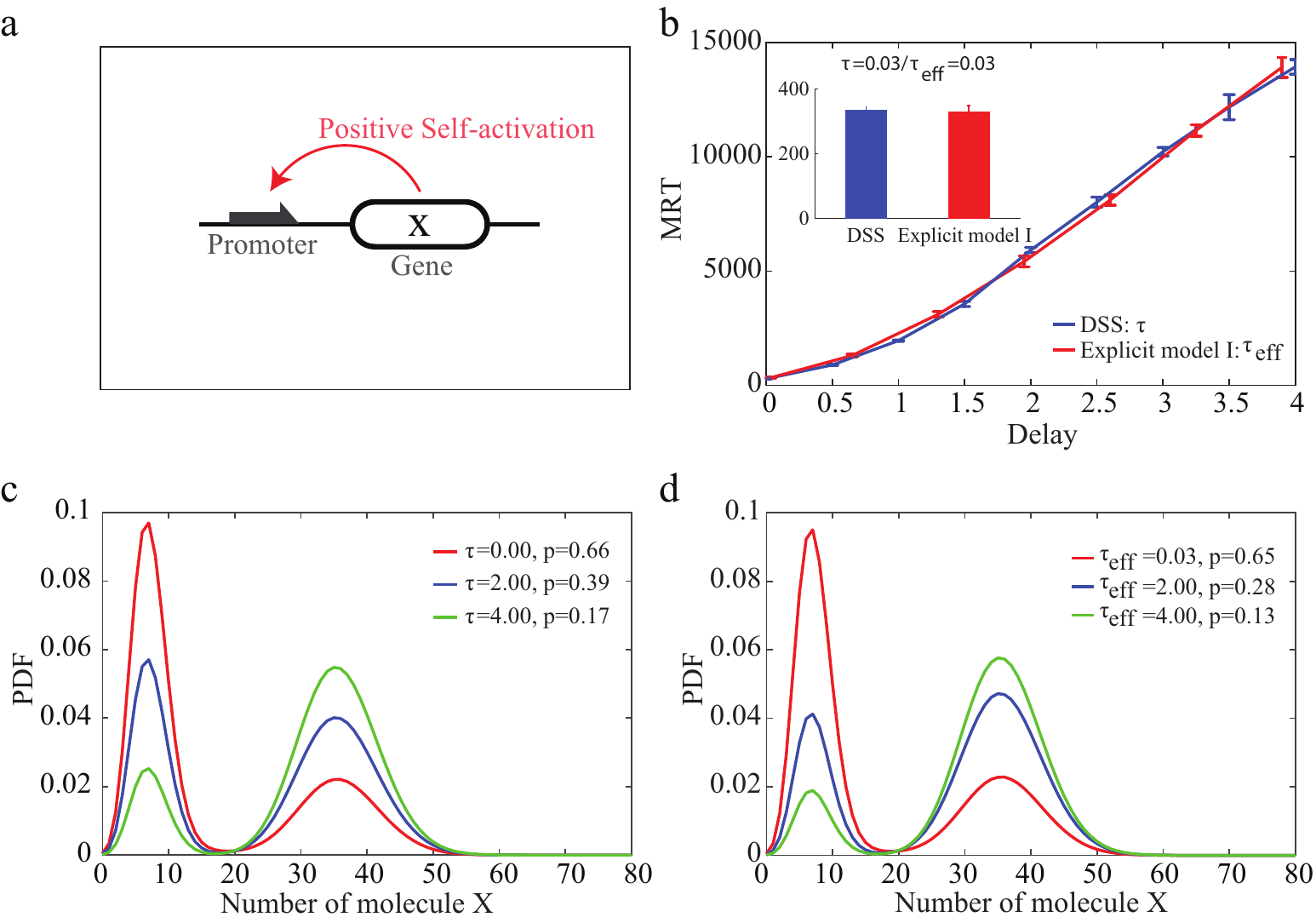}
\caption{a) A schematic diagram of self-activation circuit. b) Mean residence
time calculated with DSS and explicit model I. Inset: Mean residence time of low number state when the delay is absent. c) Equilibrium distribution
calculated with DSS. d) Equilibrium distribution calculated by explicit
model I. Legend: p is the proportion occupied by low number state. For all plots parameter values $\alpha=5,c=19,\gamma=\ln (2)$ and $b=10$ are used. MRT and PDF stand for mean residence time and probability density function respectively. }
\end{figure*}

\subsection{Explicit model I}

Suppose the delay in Eq (\ref{eq:dde1}) originates from the existence
of a precursor Y. We consider the following reaction scheme

\begin{center}
\includegraphics[width=0.45\textwidth]{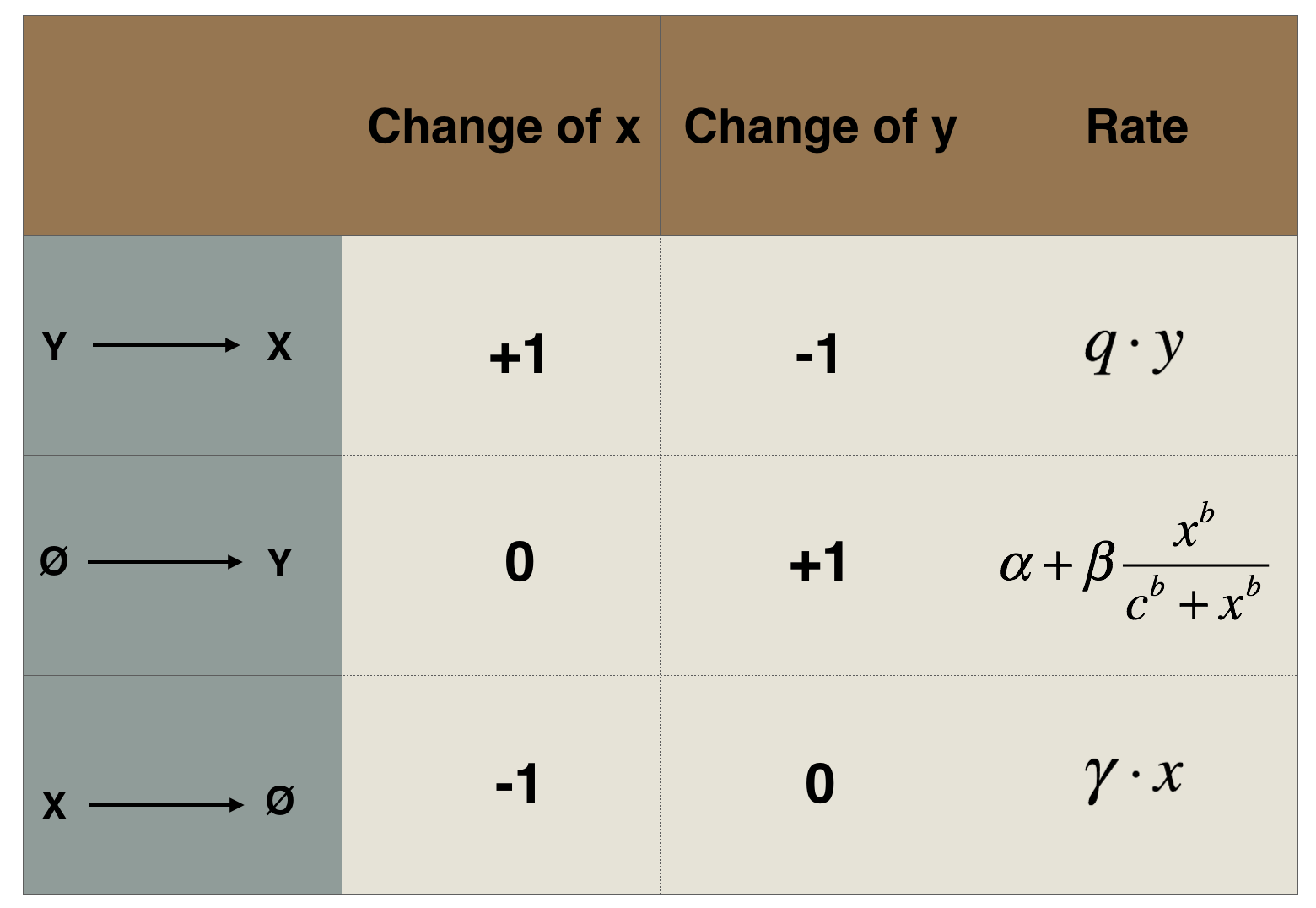}
\par\end{center}

Molecule Y is transformed into molecule X, which activates the production
of Y. At the mean field level, we can write down the corresponding ODEs to match
parameter values so as to obtain the same average value of molecules
$X$ and $Y$ given by $x$ and $y$ respectively as
\begin{equation}
\dot{x}=q\cdot y-\gamma\cdot x\label{eq:ex-x}
\end{equation}
\begin{equation}
\dot{y}=\alpha+\beta\frac{x^{b}}{c^{b}+x^{b}}-q\cdot y\label{eq:ex-y}
\end{equation}
The transformation rate $q$ sets the delay time of the system and Eq (\ref{eq:ex-x}) \& (\ref{eq:ex-y}) have the same steady states in
$x$ as in Eq (\ref{eq:dde1}) for all $q$ with all shared parameter values staying constant.

To understand the relationship between $q$ and $\tau$ we  conduct stochastic simulations of both the original DSS and the explicit process. We can tune the delay that arises from the existence of precursor by varying $q$ and adjust its value based on our simulation results. As expected, we find the delay
of the system should be proportional to $1/q$. When
the effective delay is set as $\tau_{eff}=\frac{2}{3}\cdot\frac{1}{q}$,
the mean residence time (MRT) versus $\tau_{eff}$ curve almost perfectly
collapses with the MRT versus $\tau$ calculated from the DSS (Fig.1b).
We further calculated equilibrium configurations
of the system with $\tau_{eff}=0.03,\;2.00,\;4.00$. The stationary
distribution for the explicit model is again reasonably consistent with those calculated
with DSS (Fig.1c\&1d), there being only a modest difference at $\tau_{eff}=2.00$. 

\subsection{Explicit model II}

If we regard X as a type of protein and Y as its mRNA instead of a precursor,
we can obtain a different explicit model for the same DDEs(\ref{eq:dde1}). Consider the following reactions

\begin{center}
\includegraphics[width=0.45\textwidth]{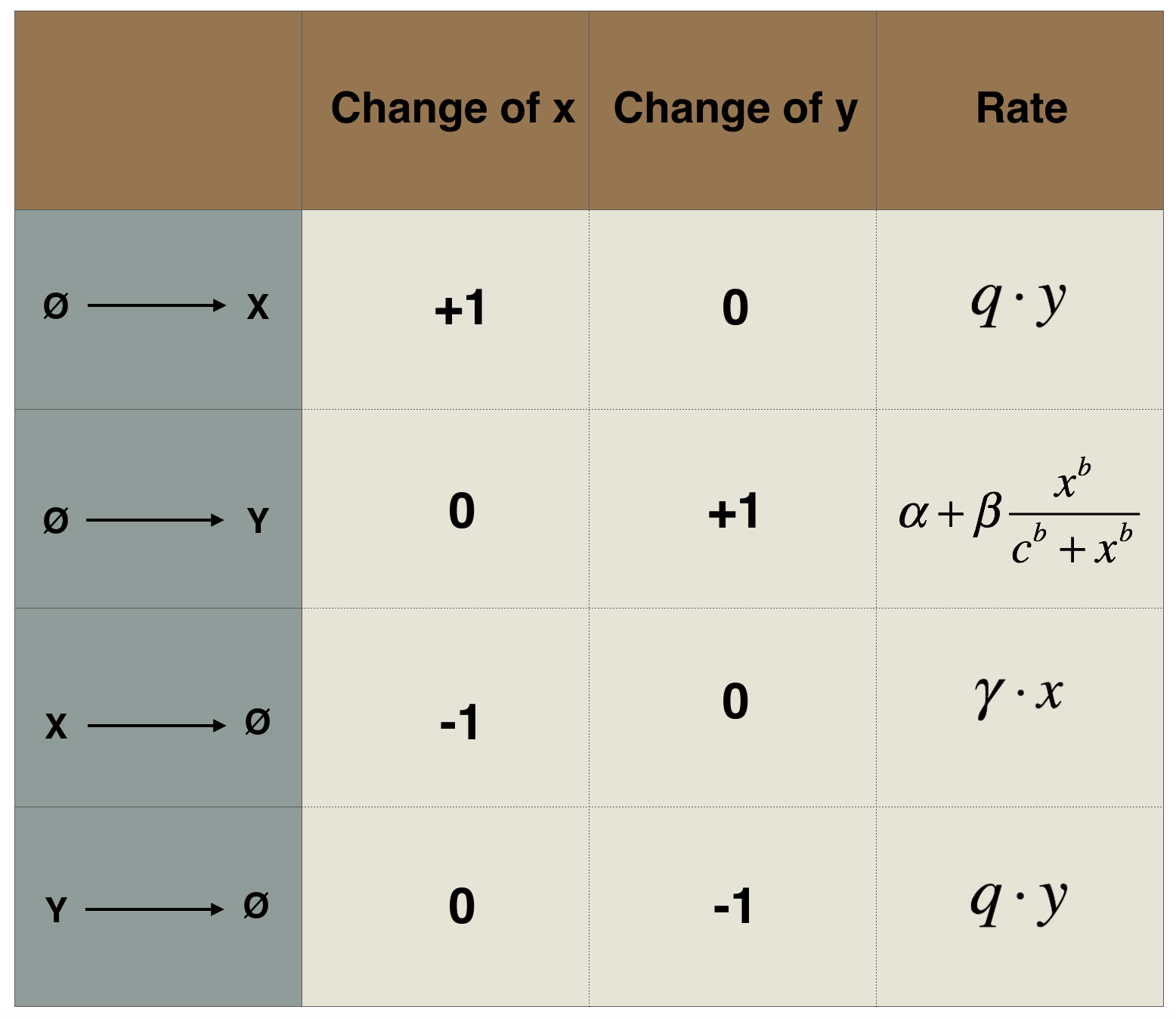}
\par\end{center}

This case is different from the precursor transformation
previously considered in that Y participates in the translation of
protein X but has an independent decay process. In contrast to
transformation, the translation process does not consume X. We have set
the decay rate of Y equal to $q$ so that the corresponding ODEs are
also identical to Eq (2)\&(3). Despite obeying the same ODEs, there
are profound differences in the MRT versus delay $\tau_{eff}$ curve and
equilibrium distribution obtained by explicit stochastic simulation. Note that we have used here the same definition for $\tau_{eff}$ as in explicit model I, but the difference in the curves cannot be accommodated by just shifting this relationship.The MRT
becomes notably smaller and even in the small delay limit ($\tau_{eff}\rightarrow0$),
the MRT does not equal to the case $\tau=0$ in the DSS (Fig.2a). Moreover, the
equilibrium distribution of the explicit model II is quantitively different from its counterpart in explicit
model I (Fig.2b). 
\begin{figure*}
\centering
\includegraphics[width=0.80\textwidth]{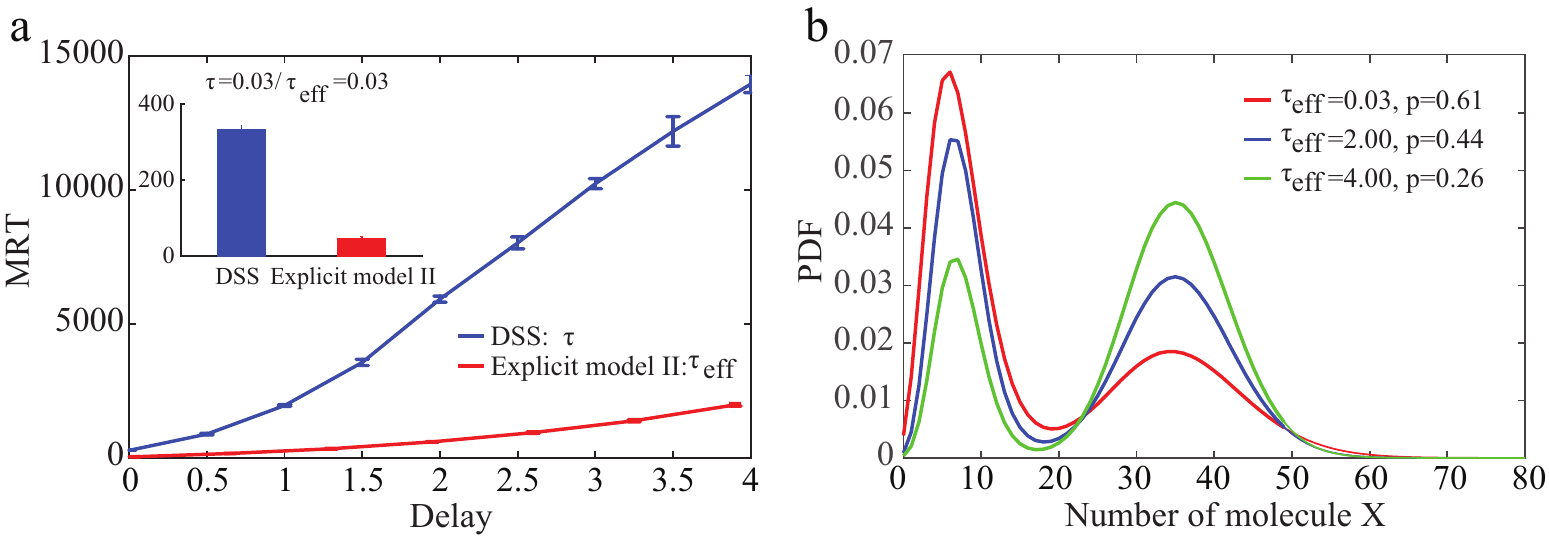}
\caption{a) Mean residence time calculated by the DSS and by explicit model II. Inset:  Mean residence time of low number state when the delay is near zero.b) Equilibrium
distribution calculated by explicit model II. Legend: p is the proportion occupied by low number state. All parameters are the same as in Fig 1.}
\end{figure*}

It is straightforward to understand the qualitative difference between explicit model I and II. Suppose at some time point $t$, the number of molecules Y happens to be
higher than the number in the steady state, due to a fluctuation. In explicit model I, such an abundant Y will quickly be transformed into X. In contrast, the production of X does not consume Y in explicit model II. Consequently, those abundant Y's produce a burst of X before they undergo independent decay. The strengthened noise in explicit model II results in the greatly reduced mean residence time. This dichotomy points out an important ambiguity in the formulation of the delay equation. In the DDE (no fluctuation) limit, these models give rise to exactly the same steady states, and there is no obvious way to choose which explicit model is better without postulating the actual delay process being modeled. Once we include stochasticity, our DSS algorithm effectively assumes that a particle placed in the queue will be transformed to X after a fixed delay (and at that time point increase X to X+1) while disappearing. This clearly is analogous to the process described by the first explicit model, which therefore agrees much more quantitatively with the original DSS.

\section{Fixed versus stochastic delay time}

Given the reasonable agreement between the explicit model I and the original DSS, we investigate in more detail the relationship between these two formulations. Let us first start with the deterministic limit given by the respective ODE systems. Starting from equation (\ref{eq:ex-y}) and given
$x(t)$, the solution of $y(t)$ is determined
as
\[
y(t)=e^{-qt}y(0)+\int_{0}^{t}ds[e^{-q(t-s)}(\alpha+\beta\frac{x(s)^{b}}{c^{b}+x(s)^{b}})]
\]
By integrating from the infinite past the initial
condition becomes negligible and we rewrite the equation above as,
\[
y(t)= \alpha/q + \int_{-\infty}^{t}ds \left[ e^{-q(t-s)}\beta\frac{x(s)^{b}}{c^{b}+x(s)^{b}} \right]
\]
Plugging back into Eq (\ref{eq:dde1}) yields
\[
\dot{x}= \alpha + \int_{-\infty}^{t}ds \left[ qe^{-q(t-s)}\beta\frac{x(s)^{b}}{c^{b}+x(s)^{b}} \right]-\gamma x
\]

From the equation above, it is clear that the delay caused by the additional
variable $ y$ follows an exponential distribution with average value
$1/q$. When $q\rightarrow\infty$, the peak of this distribution
approaches infinity and the width of the peak approaches zero. Of course, by
substituting it with a delta function distribution, we recover Eq
(\ref{eq:dde1}). The difference between the two models is that in the DDE the delay is fixed but in the explicit model the delay is exponentially distributed.

It is critical to realize that this observation regarding the difference between the two models also holds for the stochastic version. As already mentioned, one can think of the delayed reaction in the DSS algorithm as putting a produced particle into a queue and only at a fixed later time allowing it to be counted as an increase in X. The stochastic version of the explicit model creates a Y particle which then obeys a single exponential decay process to produce X; everything is the same except that the delay is now stochastic. The fact that the mean equations and the actual stochastic processes have the same relationship to each other is ultimately due to the linearity of the reaction scheme governing the production and decay of X in the explicit model.

We can now extend our notion of an explicit model to allow for more than one precursor step. For example, let us imaging that there are two precursors. 
The ODEs for the explicit models with two intermediate steps are:
\begin{equation}
\dot{x}=q\cdot z-\gamma\cdot x\label{eq:x}
\end{equation}
\begin{equation}
\dot{y}\dot{=\alpha+\beta\frac{x^{b}}{c^{b}+x^{b}}-q\cdot y}\label{eq:y}
\end{equation}
\begin{equation}
\dot{z}=q\cdot y-q\cdot z\label{eq:z}
\end{equation}
Here the molecules Y, Z are intermediate products. Assuming we know $x(t)$,
then from Eq (\ref{eq:y}),
\[
y(t)=\int_{-\infty}^{t}ds[e^{-q(t-s)}(\alpha+\beta\frac{x(s)^{b}}{c^{b}+x(s)^{b}})]
\]
Plugging it into Eq (\ref{eq:z}),
\[
z(t)=\int_{-\infty}^{t}dre^{-q(t-r)}q\int_{-\infty}^{r}dse^{-q(r-s)}(\alpha+\beta\frac{x(s)^{b}}{c^{b}+x(s)^{b}})
\]
Finally, Eq (\ref{eq:x}) becomes
\[
\dot{x}=\int_{-\infty}^{t}dr\int_{-\infty}^{r}ds[q^{2}e^{-q(t-s)}(\alpha+\beta\frac{x(s)^{b}}{c^{b}+x(s)^{b}})]-\gamma\cdot x
\]
Integrating over r first, this becomes
\[
\dot{x}=\int_{-\infty}^{t}ds[(t-s)q^{2}e^{-q(t-s)}(\alpha+\beta\frac{x(s)^{b}}{c^{b}+x(s)^{b}})]-\gamma\cdot x
\]
After some rearrangement we obtain
\[
\dot{x}=\alpha + \int_{0}^{\infty}ds'\left[ s'q^{2}e^{-qs'}\beta\frac{x(t-s')^{b}}{c^{b}+x(t-s')^{b}} \right]-\gamma\cdot x
\] 

So, the exponential distribution has been replaced by the Gamma distribution $p_2=tq^{2}e^{-qt}$. Again this holds also for the single particle stochastic dynamics where this distribution is now interpreted as the time it takes for a particle to be transformed from Y $\rightarrow$ Z $\rightarrow$ X, where each of the reactions is irreversible and occurs at the same rate $p$. A simple extension of the above shows that 
$$
p_{n}(t)=\frac{n^{n}t^{n-1}}{\tau^{n}(n-1)!}e^{-\frac{n}{\tau}t}
$$
where now we have defined $\tau = n/q$ This can be proven by induction, using 
$p_{n}(t)=\int_{0}^{t}p_{n-1}(t')p_{1}(t-t')dt'=\frac{q^{n}t^{n-1}}{(n-1)!}e^{-qt}$. When $t^{*}=\frac{n-1}{n} \tau $, $p_{n}(t)$
reaches a maximum. As we vary the number of intermediate steps $n$
and keep the mean value of delay $<t> = \tau$ the same, the distribution becomes increasingly sharp. A plot of $p_{n}(t)$ is shown in Fig. 3. 

Hence, the limiting process of making $n$ large leads to a precise fixed value of the delay and asymptotically approaches the DSS. It then becomes a quantitative issue as to whether the actual process has intermediate states and to what extent they occur at roughly equal rates, as opposed to having one step dominate (being rate-limiting), an whether the fixed delay version is a good enough approximation for that actual situation. For the simple self-activation case, we have shown that even with only one precursor the DSS is a reasonably accurate approach.
\begin{figure}
\centering
\includegraphics[width=0.45\textwidth]{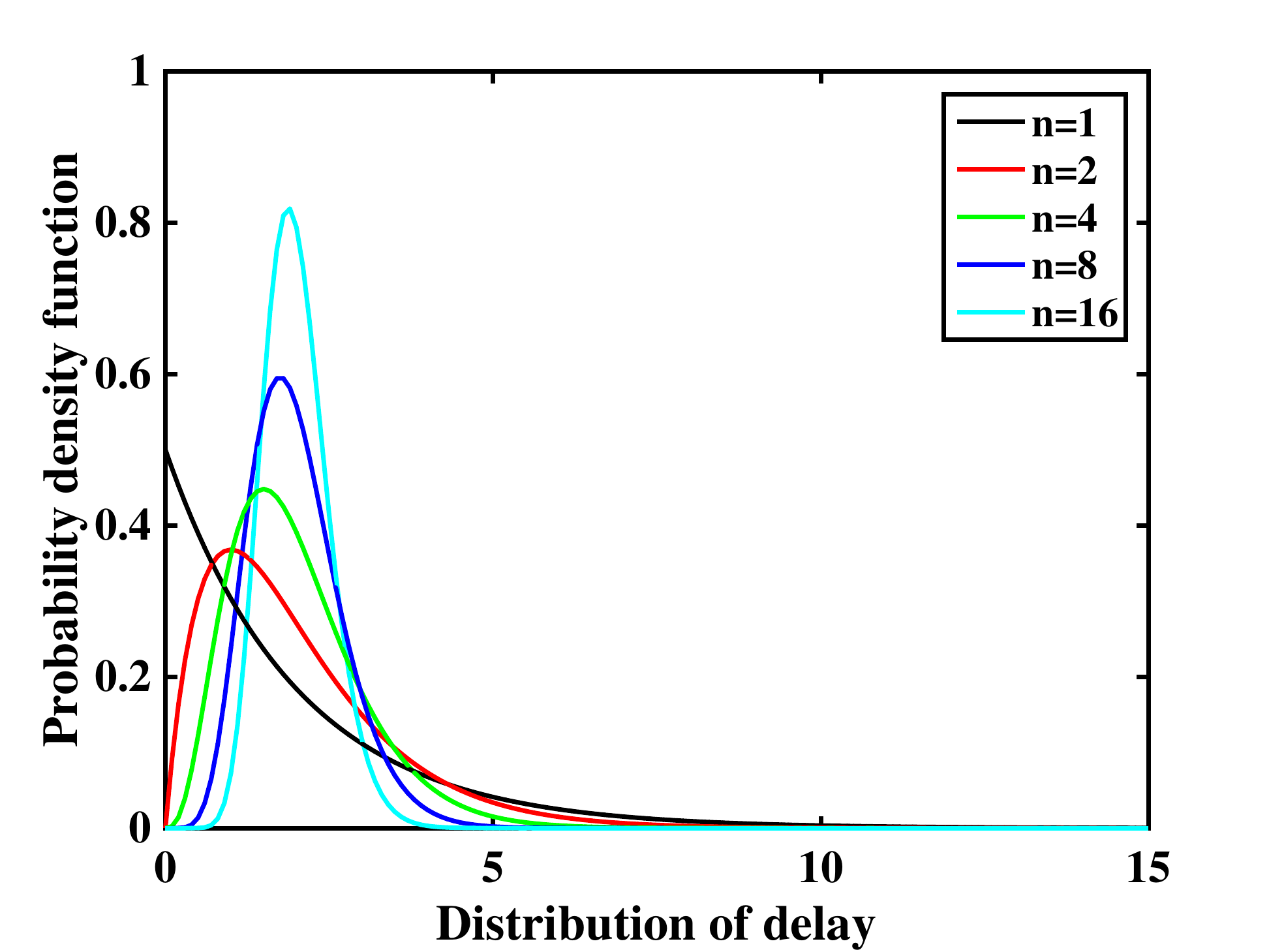}
\caption{Probability density function (PDF) $p_{n}$. Here the mean delay is fixed to be 2. }
\end{figure}

\section{The toggle switch}

We now extend our discussion to a more complex circuit, the toggle switch shown schematically in Fig.4a. If the average number
of molecules X and molecule Y are represented by $x$ and $y$, then
the time evolution of $x$ and $y$ is determined by the following
DDEs, (to simplify the problem, we have assumed that the delay exists only in
the repressive regulation from Y to X. )
\begin{equation}
\dot{x}=\beta\frac{1}{1+y(t-\tau)^{2}/K^2}-\gamma\cdot x
\end{equation}
\begin{equation}
\dot{y}=\beta\frac{1}{1+x^{2}/K^2}-\gamma\cdot y
\end{equation}
where $\beta$ is the decrease in transcription rate due to protein
binding to the promoter, $K$ the concentration of
X and Y needed for half-maximal reduction, $\gamma$ the degradation
rate coefficient of the protein, and $\tau$ the transcriptional delay
time. This DDE is again extended to a DSS by using the rates on the right hand side of the above equations. We have chosen to use the same parameters as in \cite{key-2},
which puts the system in a bistable regime. Similar
to the result for the self-activation
circuit, the mean residence time of the $X<Y$ state grows rapidly as
delay increases (Fig.4b). The equilibrium distribution does not 
change significantly with varying delay and the probability of finding
molecule levels in the attractive basin of each stable state are approximately equal (Fig.4c). 
\begin{figure*}
\centering
\includegraphics[width=0.80\textwidth]{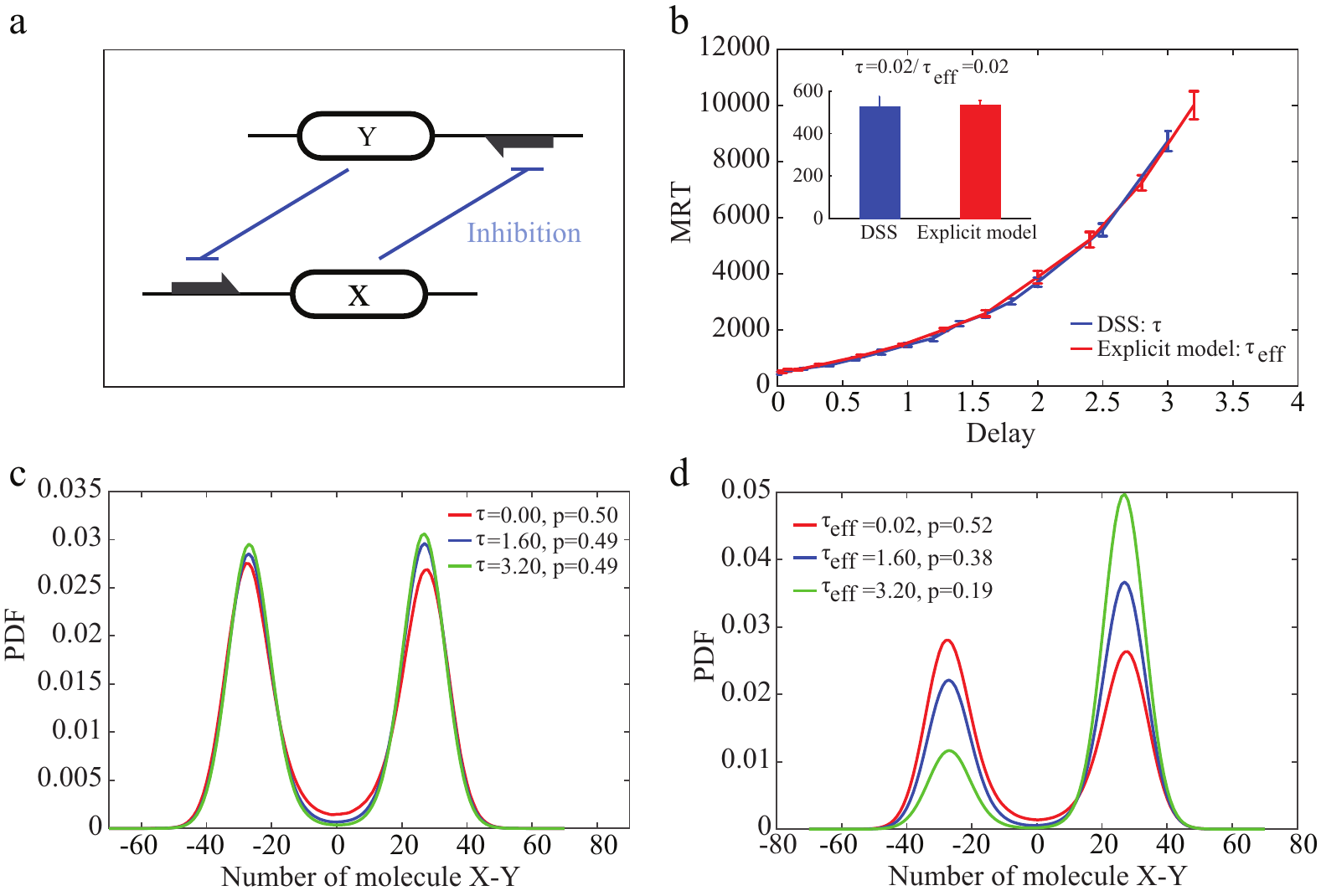}
\caption{a) A schematic diagram of toggle-switch circuit. b) Mean residence
time of $X<Y$ state calculated by DSSs and the explicit model. Inset:  Mean residence time when the delay is absent. c) Equilibrium distribution
calculated by DSSs. d) Equilibrium distribution calculated by explicit
model. Legend: $P_0$ is the proportion occupied by the $X<Y$ number state. $\beta=21.9, k=6.8, \gamma=\ln (2)$}
\end{figure*}

We now construct the related explicit model, assuming that the delay in Eq (4)\&(5) originates from
the existence of a precursor Z. We consider the following reactions,

\begin{center}
\includegraphics[width=0.45\textwidth]{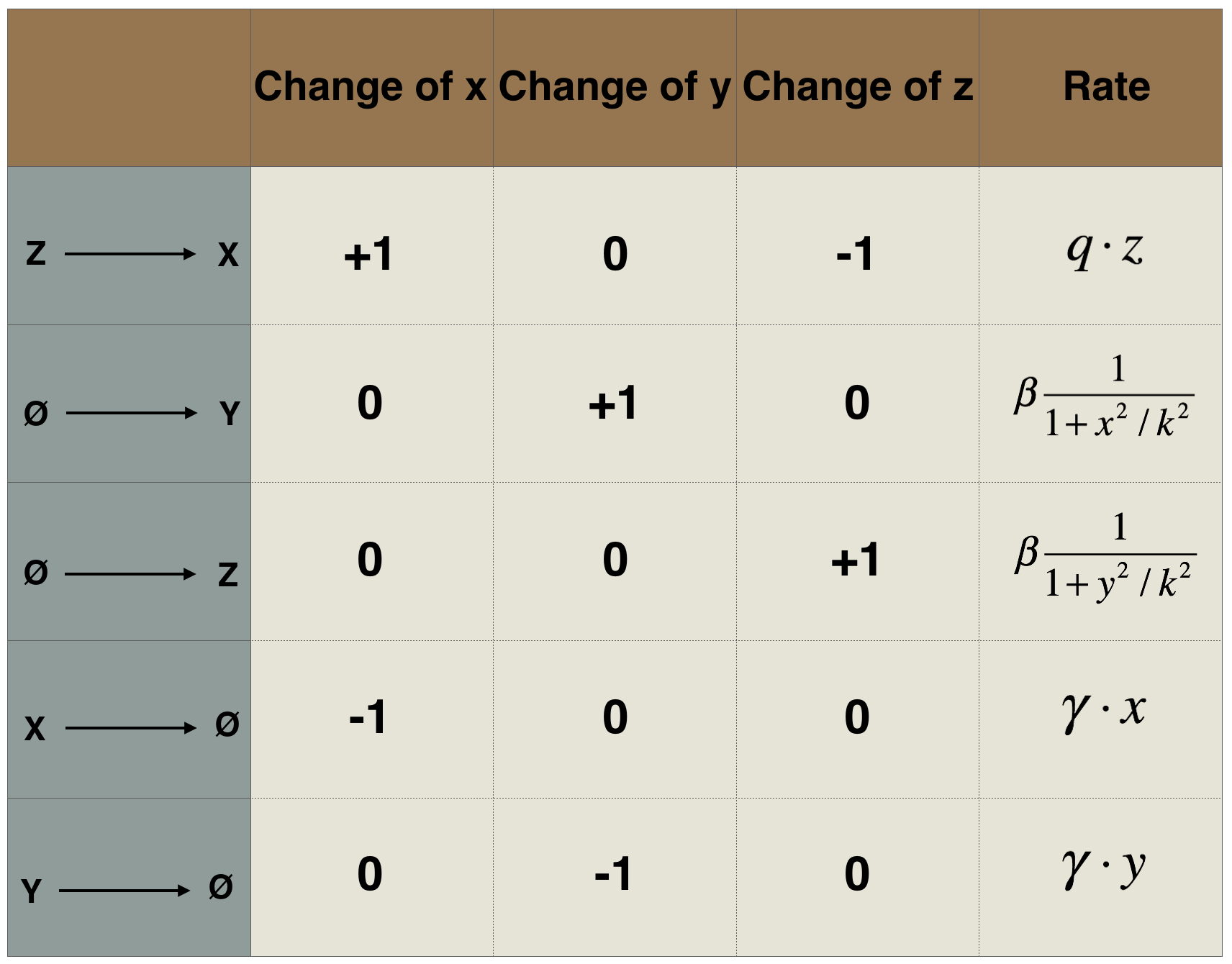}
\par\end{center}

Molecule Z is transformed into molecule X, which is a repressor of
Y. Molecule Y further inhibits the production of Z. The corresponding
ODEs are
\begin{equation}
\dot{x}=q\cdot z-\gamma\cdot x
\end{equation}
\begin{equation}
\dot{y}=\beta\frac{1}{1+x^{2}/k}-\gamma\cdot y
\end{equation}
\begin{equation}
\dot{z}=\beta\frac{1}{1+y^{2}/k}-q\cdot z
\end{equation}
By construction, Eq (9) - (11) have same steady states of $x$ as in Eq (7)\&(8).

We can tune the delay that arises from the existence of a precursor by
varying the $q$ value. The delay of the system is proportional to
$1/q$ in the same manner as we have seen in the self-activation circuit. When
the effective delay is defined as $\tau_{eff}=0.80\cdot\frac{1}{q}$,
we find that the MRT of $X<Y$ state versus $\tau_{eff}$ curve almost perfectly
collapses with MRT versus $\tau$ calculated from SDDEs (Fig.4b). However,
the equilibrium distribution in this explicit model is strongly influenced
by the value of the delay, which suggests that the MRT of $X>Y$ state versus $\tau_{eff}$
curve does not agree with its counterpart in the DSS. Alternatively, one could get a better match to the decay of the $X>Y$ state and fail to match this one (data not shown). This is in stark
contrast to the delay-independent equilibrium distribution in the DSS
(Fig.4c \& 4d) which shows no such change. 

As discussed above, the DSS results should be approached asymptotically if the number of intermediate states is increased. 
\begin{figure}[!h]
\centering
\includegraphics[width=0.45\textwidth]{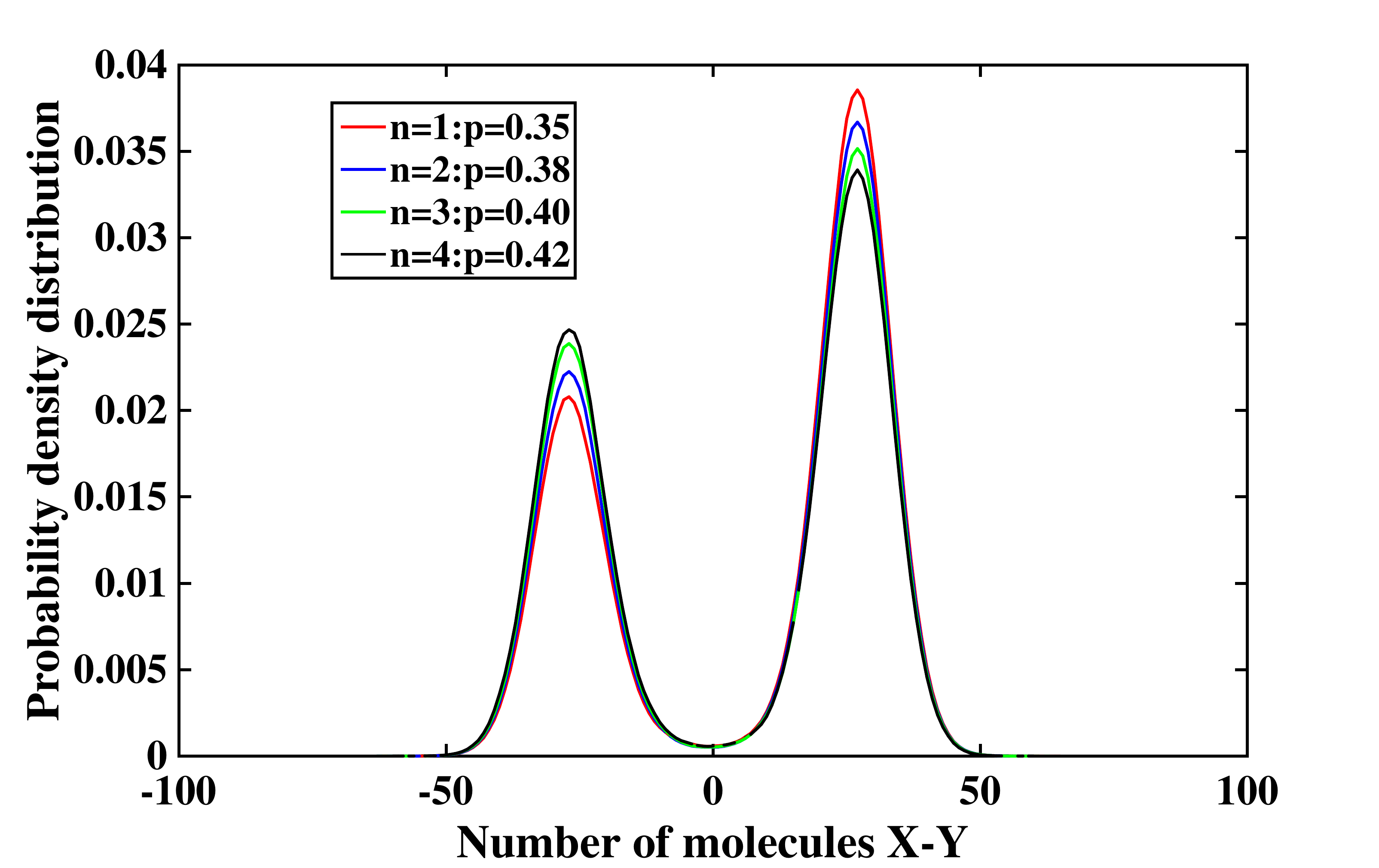}
\caption{Equilibrium distribution of toggle-switch circuit for $\tau=2$. The
inset shows the probability of $X<Y$. Recall that in the DSS, $P(X<Y)\sim0.5$. }
\end{figure}
We test the rapidity of this convergence in Fig. 5. As we increase the number of intermediate reactions $n$, the difference in the height of two peaks becomes smaller, as expected. Yet the difference is not negligible even for the relatively large number of intermediate reactions, $n = 4$. The width of the delay time distribution is still fairly significant at $n=4$  (Fig. 3). Apparently, the extra nonlinearity in the toggle switch circuit makes the system more sensitive to having such a non-trivial distribution.

\section{Delay-induced oscillation}

Previous studies have argued that the introduction
of delay in otherwise stable systems can induce oscillations \citep{key-1,key-3,key-4,key-10,key-11,key-12,key-13,key-14,key-15}.
Here we focus on the case of delayed protein decay, which has been
shown to undergo oscillations in a DSS formulation \citep{key-3}. Furthermore, it has been posited that this oscillation can be partially understood by writing down the DDE system for average number of protein X, represented by $x$, as 
\begin{equation}
\frac{dx}{dt}=A-B\cdot x(t)-C\cdot x(t-\tau )\label{eq:oscillation}
\end{equation}
where A is the rate of protein production, and $B,\;C$ the rates of
non-delayed and delayed degradation respectively. Here we show that both of these statements need to carefully reconsidered.

First, it is necessary to note that there is an inherent ambiguity in how to define the DSS for this case. We need to specify in particular whether a particle slated for a delayed decay can undergo regular decay while waiting in the queue. A master equation formulation of the stochastic version of Eq. \ref{eq:oscillation} seems to allow this  to occur (see Ref. \cite{key-3}), but for the parameter set reported in that work the
characteristic direct decay time $1/B$ is much smaller than the delay
$\tau$ and therefore nearly all molecules X involved
in delayed decay (i.e. placed in the queue waiting to decay) cannot finish this process and undergo direct decay
instead. As a consequence, the last term on the right side of Eq (\ref{eq:oscillation})
would not play any role in a stochastic simulation. 

Consequently, in our simulation we prohibit molecules
undergoing delayed decay from participating in direct decay. With the same parameter set used in \citep{key-3}, $x$
oscillates (Fig.6a). The power spectrum
calculated from time series of $x$ (Fig.6c) reveals oscillatory
behavior by the location of the peaks. As expected these are separated by $1/\tau$. But, it is clear that the system is not accurately described by the above equation, even in an average sense. The simplest way to see this is to note that the mean value of X depends on the delay, whereas the stead-state solution of the equation does not. The fact that this equation can have oscillatory modes cannot be relevant for whether or not the stochastic system oscillates. 

We now construct an explicit model analog of our DSS. Protein degradation often occurs through a sequence of events that
are mediated by a complex proteolytic pathway \cite{key-7}. It is
thus reasonable to assume in the delayed degradation reaction, protein
X will first be transformed into an intermediate product Y, which
has an independent decay process. The existence of the intermediate product
Y causes the delay in the degradation of X \cite{key-7}. Here are
the reactions involved, 
\begin{center}
\includegraphics[width=0.45\textwidth]{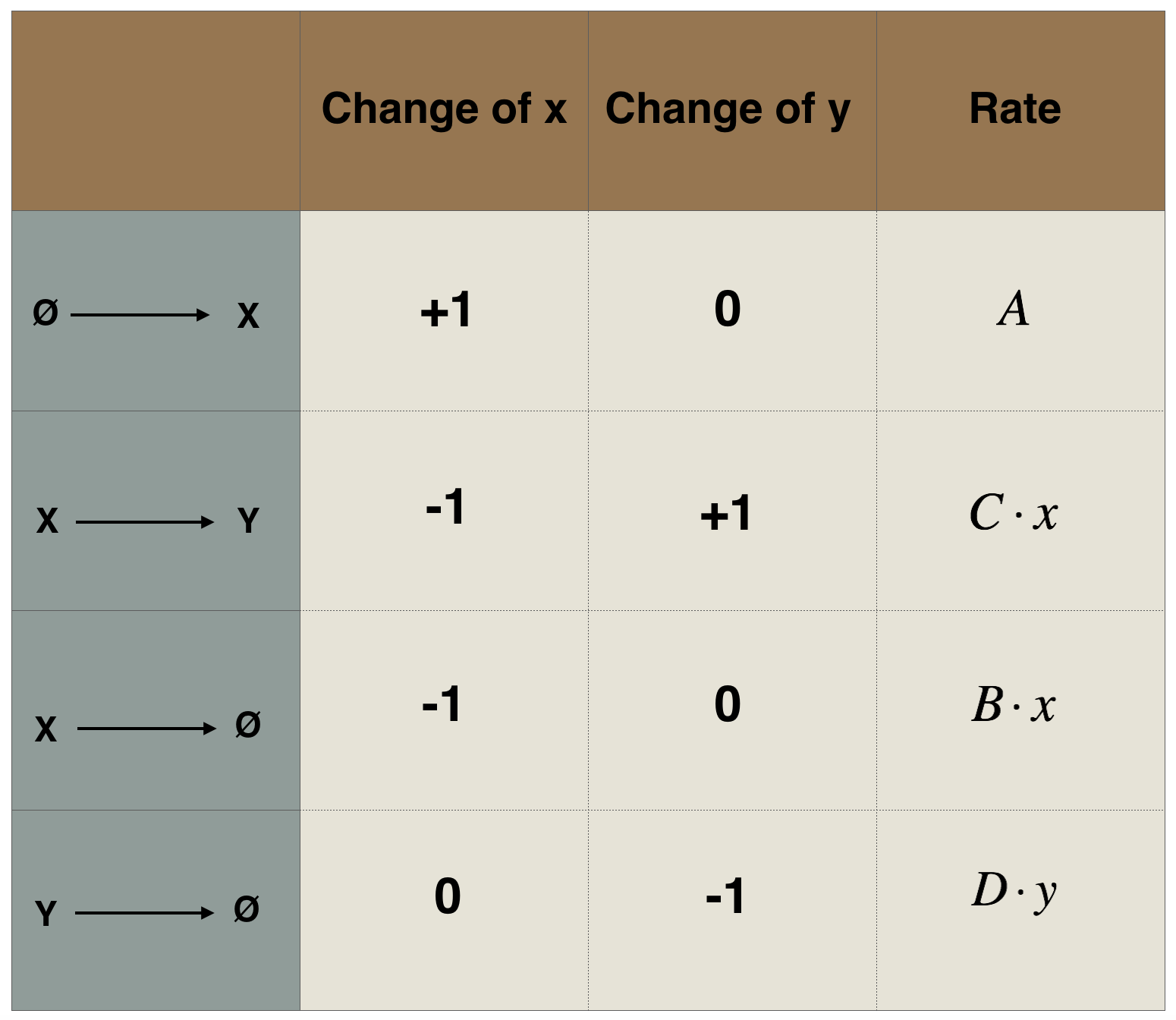}
\end{center}
The corresponding ODEs in the deterministic limit are are
\begin{equation}
\frac{dx}{dt}=A-B\cdot x-C\cdot x
\end{equation}
\begin{equation}
\frac{dy}{dt}=C\cdot x-D\cdot y
\end{equation}
The average value of delay is $1/D$. Therefore we set $D=1/\tau$
in our explicit model to match the DSS. Note that unlike the previous deterministic equation, the steady-state value of the total number of particles $(x+y)$ does depend on $D$; it equals $\frac{A}{B+C} \left( 1+ C \tau \right)$ which scales linearly for long time delay and agrees with the data in Fig. 6a.

\begin{figure*}
\centering
\includegraphics[width=0.80\textwidth]{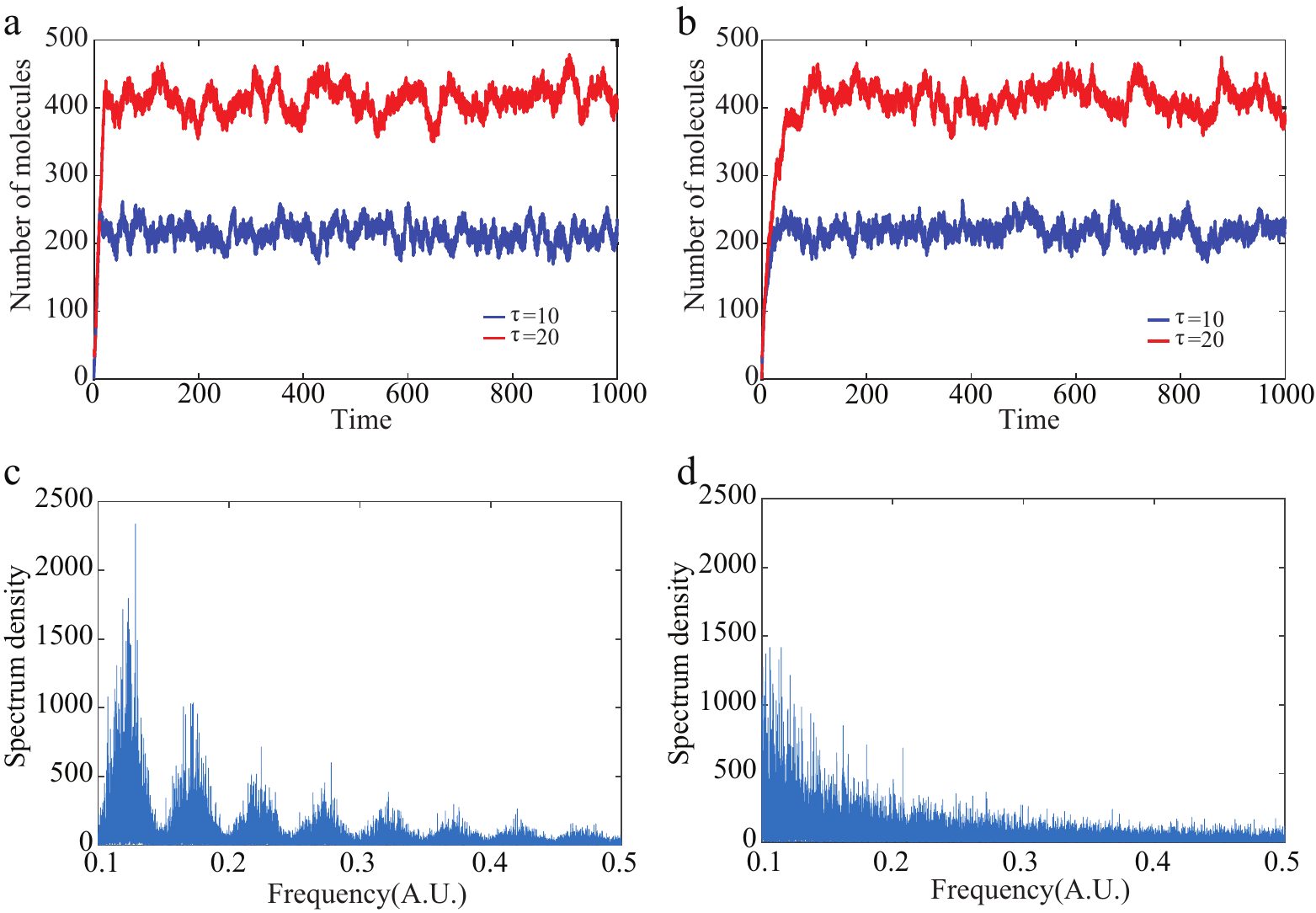}
\caption{a)Time series of total number of molecules in SDDE. b)Time series
of total number of molecules in explicit model. c)Corresponding periodogram
in SDDE. d) Corresponding periodogram in explicit model. Here $A=100$,
$B=4.1$, $C=1.0$, $D=1/\tau$.}
\end{figure*}

For the case of linear reactions there can be no oscillations at the deterministic level. Since the system is linear, oscillations must mean imaginary eigenvalues of the Jacobian matrix
 \begin{displaymath} 
\mathbf{J} = \left( \begin{array}{ccc} 
-B-C & 0  \\ 
C & -D  
\end{array} \right) 
\end{displaymath}
A simple calculation shows however that the eigenvalues are $-B-C$ and $-D$, yielding simple exponential relaxation. In fact it is trivial to extend this result to the case of an arbitrary number of intermediates each of which is produced and decays via unimolecular reactions.  In other words, the exact solution
of any explicit model predicts no oscillatory behavior in the mean field limit. Any oscillations must be due to stochasticity.

In Fig 6b we show a simulated time series for the total particle number in a one intermediate explicit model, and its power spectrum is presented in Fig. 6d.
The time series of $X$ generated by the DSS versus the explicit model look superficially similar (Fig.6a\&6b); however,
the power spectrum of SDDE and explicit model are markedly different.
In contrast to the equally spaced peaks in the power spectrum (Fig. 6c), there is no obvious
peak in the explicit model (Fig.6d). Thus, the exponential distribution of delay values will wash out the oscillation. We have extended this calculation to the case of $n=4$ (Fig.7) which has a somewhat peaked delay distribution. Even here though, spectral peaks cannot be detected as the distribution is still wide enough to eliminate the peaks related to the fixed delay.
\begin{figure}
\includegraphics[width=0.40\textwidth]{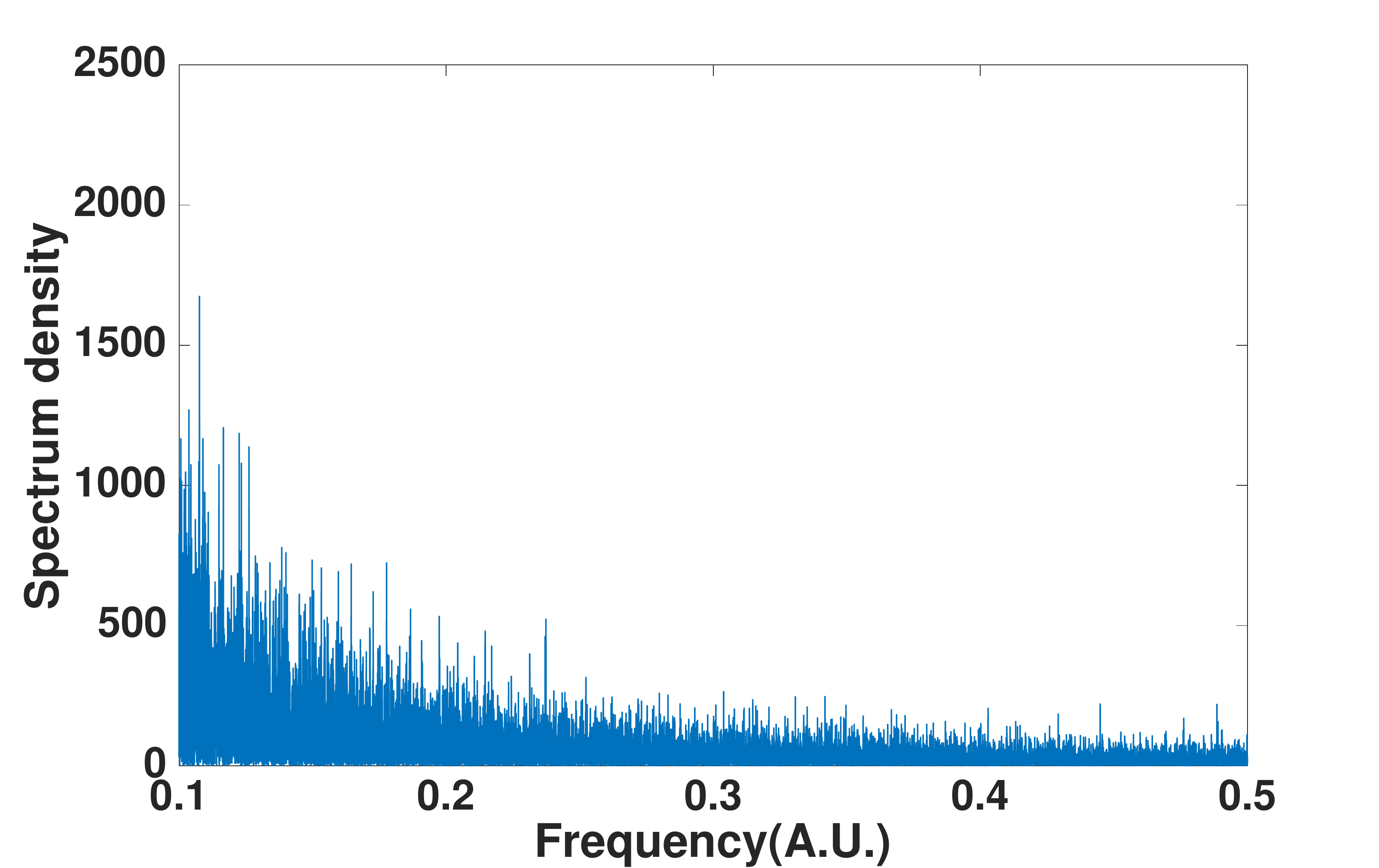}
\caption{Time series
of total number of molecules in explicit model with $n=4$. Parameters are the same as in Fig. 6. }
\end{figure}

The results here and in the previous section on the toggle switch address the importance of delay
distributions. Our results show that 
even when the number of intermediate reactions is increased up to
four, there can still be non-negligible differences 
between DSSs and explicit models. Modeling of biological systems may require constructing explicit systems if one wants to obtain quantitatively accurate predictions. 

\section{Discussion}

Stochastic delayed differential systems have been very popular in biological physics
due to their relative simplicity as compared to models that include a large number of intermediate steps that are anyway not being monitored in the experimental data. The cost of such
simplicity is the conversion from Markovian explicit models to non-Markovian
DSSs. In most cases, the non-Markovian property makes analytical studies challenging\cite{key-28, key-29, key-30, key-31}. When the delay is much larger
than the transition time between stable states, it can be assumed
either the delay does not affect the dynamics within each attractive
basin or the joint probability $P(X(t),X(t-\tau)$ can be decoupled
as $P(X(t))\cdot P(X(t-\tau))$. Approximate analytical solutions
can be derived with such assumptions\citep{key-15,key-3}. In the
small delay case, it is sometimes possible to derive approximate solutions
for simple cases\citep{key-16}. As for moderate delay problems,
to the best of our knowledge, there is no good way to derive analytical
solutions, even approximately. 

Because of the difficulty in solving DSSs analytically, two different
but consistent stochastic simulation methods have been proposed to
study these systems numerically\citep{key-3,key-17,key-18}. Since the reaction
rate depends on both $X(t)$ and $X(t-\tau)$, both methods require
the storage of system dynamics from $t$ to $t-\tau$. Therefore stochastic
simulation methods become computationally inefficient for large $\tau$.
textcolor{blue}{As we have seen in our examples the rates of intermediate reactions in explicit models are proportional to $1/\tau$, so that long delays correspond to slow reactions.  However, slow reactions do not increase the computational cost of a stochastic simulation.Thus for systems with long delays explicit models may be computationally preferable.}

Beyond the issue of computational ease is the question of quantitative reliability.
In this paper, we have demonstrated that DDEs often yield inaccurate
transition times and equilibrium distributions.  Additionally, there can exist multiple explicit models with fundamentally different dynamics that give rise to the same DDEs; some of these have stochastic extensions which correspond better than others to a given DSS; sometimes non-uniqueness exists when we
attempt to formulate stochastic simulation directly to DDEs, as we have seen in the delay-induced oscillation case. Consequently results that depend strongly on having a fixed delay may be non-robust when the cause of the delay is handled explicitly. In the end, we argue that more attention needs to be paid to the limitation of the DSS approach; blind use of this approach may cause significant mischaracterization
of important biological systems. 

\bibliography{delay}

\begin{thebibliography}{10}
\bibitem{key-20}
Residence time is defined as the time duration the system stays in the
  attractive basin of a stable state. it is also called first passage time or
  transition time.
\end{thebibliography}
\bibliographystyle{unsrtnat}

\end{document}